\documentclass[prl,aps,superscriptaddress,twocolumn,floatfix,10pt]{revtex4-2}

\usepackage{ragged2e}

\usepackage{dsfont}

\usepackage{mathtools}
\usepackage{amsmath}
\usepackage{amssymb,mathrsfs}
\usepackage{graphicx}
\usepackage{epstopdf}
\usepackage{mathtext}
\usepackage{yfonts}
\usepackage{bm}

\usepackage{soul}

\usepackage{color}
\usepackage[colorlinks=true,linkcolor=blue]{hyperref}

\begin{document}

\title{Observation of topological vortex solitons on disclinations}

\author{A.~V.~Kireev}
\affiliation{Institute of Spectroscopy, Russian Academy of Sciences, Troitsk, Moscow, 108840, Russia}
\affiliation{Moscow Institute of Physics and Technology (National Research University), Dolgoprudny, 141701, Russia}

\author{K.~Sabour}
\affiliation{Moscow Institute of Physics and Technology (National Research University), Dolgoprudny, 141701, Russia}

\author{V.~O.~Kompanets}
\affiliation{Institute of Spectroscopy, Russian Academy of Sciences, Troitsk, Moscow, 108840, Russia}

\author{S.~Y.~Alyatkin}
\affiliation{Hybrid Photonics Laboratory, Skolkovo Institute of Science and Technology, Territory of Innovation Center Skolkovo, Bolshoy Boulevard 30, building 1, Moscow, 121205, Russia}

\author{N.~S.~Kostyuchenko}
\affiliation{Institute of Spectroscopy, Russian Academy of Sciences, Troitsk, Moscow, 108840, Russia}
\affiliation{Quantum Technology Centre, Faculty of Physics, M.~V.~Lomonosov Moscow State University, Moscow, 119991, Russia}

\author{S.~A.~Zhuravitskii}
\affiliation{Institute of Spectroscopy, Russian Academy of Sciences, Troitsk, Moscow, 108840, Russia}
\affiliation{Quantum Technology Centre, Faculty of Physics, M.~V.~Lomonosov Moscow State University, Moscow, 119991, Russia}

\author{N.~N.~Skryabin}
\affiliation{Quantum Technology Centre, Faculty of Physics, M.~V.~Lomonosov Moscow State University, Moscow, 119991, Russia}

\author{I.~V.~Dyakonov}
\affiliation{Quantum Technology Centre, Faculty of Physics, M.~V.~Lomonosov Moscow State University, Moscow, 119991, Russia}

\author{A.~A.~Kalinkin}
\affiliation{Quantum Technology Centre, Faculty of Physics, M.~V.~Lomonosov Moscow State University, Moscow, 119991, Russia}

\author{K.~A.~Sitnik}
\affiliation{Hybrid Photonics Laboratory, Skolkovo Institute of Science and Technology, Territory of Innovation Center Skolkovo, Bolshoy Boulevard 30, building 1, Moscow, 121205, Russia}

\author{S.~K.~Ivanov}
\affiliation{Instituto de Ciencia de los Materiales, Universidad de Valencia, Catedr\'{a}tico J. Beltr\'{a}n, 2, Paterna, 46980, Spain}

\author{S.~P.~Kulik}
\affiliation{Institute of Spectroscopy, Russian Academy of Sciences, Troitsk, Moscow, 108840, Russia}
\affiliation{Quantum Technology Centre, Faculty of Physics, M.~V.~Lomonosov Moscow State University, Moscow, 119991, Russia}

\author{S.~V.~Chekalin}
\affiliation{Institute of Spectroscopy, Russian Academy of Sciences, Troitsk, Moscow, 108840, Russia}

\author{P.~G.~Lagoudakis}
\affiliation{Hybrid Photonics Laboratory, Skolkovo Institute of Science and Technology, Territory of Innovation Center Skolkovo, Bolshoy Boulevard 30, building 1, Moscow, 121205, Russia}

\author{V.~N.~Zadkov}
\affiliation{Institute of Spectroscopy, Russian Academy of Sciences, Troitsk, Moscow, 108840, Russia}
\affiliation{Faculty of Physics, Higher School of Economics, Moscow, 105066, Russia}

\author{Y.~V.~Kartashov}
\email[Correspondence email address: ]{kartashov@isan.troitsk.ru}
\affiliation{Institute of Spectroscopy, Russian Academy of Sciences, Troitsk, Moscow, 108840, Russia}

\begin{abstract}
Vortex-carrying wave fields play a crucial role in photonics due to unusual propagation properties and interactions with matter, which enable numerous practical applications ranging from optical tweezers and imaging to information encoding and transmission. Localized vortex-carrying beams propagating in nonlinear optical media may form self-sustained excited states---vortex solitons---which are however usually prone to instabilities and require high powers for their stabilization in non-topological materials. Using fs-laser written aperiodic waveguide arrays, we demonstrate that photonic topological insulators with disclinations admit the formation of stable and thresholdless vortex solitons with tunable shapes. These unique materials belong to a class of higher-order topological insulators and allow the propagation of localized, topologically protected excitations at the disclination core, enabling disorder-resistant transmission of signals and energy. We show that vortex solitons bifurcate from the superposition of topologically protected linear edge states at the disclination core and remain stable in the entire forbidden topological gap. Realized topological vortex solitons with symmetries that are inaccessible in periodic lattices are the first example of excited soliton states with non-trivial phase structure in topological insulator. Our findings shine light on the interplay between nonlinearity, the angular momentum degree of freedom of light, and the material topology.
\end{abstract}

\maketitle


The development of state-of-the-art approaches to control the propagation of light and the internal structure of laser radiation is at the heart of modern photonics. New mechanisms of confinement and transmission of light became available with the advent of topological insulators (TIs)---unusual physical materials that allow the formation of localized states at their boundaries, a property directly related to the non-trivial topology of bulk bands. TIs, originally discovered in solid-state physics \cite{1,2}, have been realized and found diverse practical applications also in acoustics~\cite{3}, cold atoms and matter-wave physics~\cite{4,5}, polariton condensates~\cite{6,7}, metamaterial physics~\cite{8}, and photonics~\cite{9,10,11,12}. A $d$-dimensional TI possesses a forbidden gap in its bulk, just like a conventional insulator, but at the same time it supports $(d-1)$-dimensional states at its edges that are localized across the edge with energies belonging to the forbidden gap. Such states are the natural consequence and indicator of the non-trivial band topology. Once they cannot be destroyed without changing the topology of the system, they exhibit exceptional robustness to disorder and edge imperfections upon propagation. For example, in Chern or some Floquet TIs the edge states are unidirectional and traverse defects without backward reflection. A non-trivial band topology in TIs usually emerges due to breakup of certain symmetries of the system, such as time-reversal or inversion symmetry, from deformations of the unit cells or temporal modulations of the system parameters~\cite{9,10,11,12}. In particular, in the recently introduced class of higher-order topological insulators (HOTIs), topological phases can be realized by shifting sites within the unit cells of the underlying lattice, which affects intra- and intercell couplings between lattice sites. This allows to realize  $n$-th order HOTIs on $d$-dimensional lattices supporting localized topological states with an effective dimensionality $(d-n)$ that can be more than one dimension below that of the bulk, see recent reviews on electronic~\cite{13}, acoustic and photonic~\cite{14,15} HOTIs. For example, second-order 2D HOTIs support effectively 0D corner states in addition to 1D edge states, as has been shown on various photonic platforms~\cite{16,17,18,19,20,21,22}.

TIs created in optical materials enable the study of unique interplay between topology and nonlinearity, since such materials can possess a strong nonlinear response. Nonlinearity introduces desired tunability into properties of photonic TIs and opens new prospects for the observation of nonlinear phenomena that benefit from confinement of topological modes. These include generation of new frequencies and lasing in topological states, formation of topological states with new symmetry, their interactions and all-optical control of localization degree, internal structure, and evolution dynamics~\cite{23,24}. Nonlinearity in TIs produces a unique class of topological edge solitons emanating from linear edge states and inheriting from them topological protection. Depending on the type of TI, topological solitons can circulate along its edge while maintaining localization not only across, but also along the edge of TI~\cite{25,26,27,28,29,30} (nonlinearity can induce such circulation too~\cite{31,32}) or exist as immobile states at the interfaces~\cite{33,34,35} of topologically distinct materials. In turn, HOTIs support topological solitons localized in their outer corners \cite{36,37,38}. Strong and tunable localization of the latter states is particularly beneficial for enhancement of nonlinear interactions.

Nevertheless, all topological edge solitons observed so far experimentally represent the simplest fundamental states that inherit their internal structure from single linear edge state from which they bifurcate in topological gap. Experimental observation of topological solitons carrying vorticity remained elusive until now. At the same time, the very possibility to realize topological protection for high-intensity vortex-carrying states that can stably propagate in nonlinear medium, appears as a problem of fundamental significance due to the possibility to use such states for power and information transfer~\cite{39}, generation of new frequencies and waves carrying different topological charges~\cite{40}, nonlinearity-controlled switching between different vortex modes, realization of vortex lasing~\cite{41}, and many other applications.

In this respect, recently introduced topological materials with disclinations in aperiodic lattices with crystallographic defects that globally modify the lattice structure open new perspectives for the implementation of topological vortex-carrying modes, since such materials may possess new types of discrete rotational symmetry that are not accessible for periodic lattices. Topological disclinations can trap fractional spectral charges and host multiple coexisting localized topological states with different internal structure at the disclination core~\cite{42,43,44,45} located in the center of the structure, in addition to usual topological states in its outer corners~\cite{46,47}. This property is of crucial importance for the realization of vortex modes in such systems belonging to the HOTI class, because in contrast to recently observed single-site nonlinear states at the disclination core~\cite{48}, vortex solitons can result from combination of several degenerate topological states. In contrast to non-topological vortex solitons in periodic lattices~\cite{49} that are prone to azimuthal instabilities due to their excited nature and require considerable threshold power~\cite{40,50,51}, topological vortex solitons on disclinations are expected to be thresholdless and exhibit improved stability properties. Although linear vortices have been observed within a single channel of disclination lattice with multimode waveguides~\cite{52}, no vortex solitons have been reported that would occupy multiple waveguides on the disclination core, reflecting its discrete rotational symmetry and the global crystallographic deformation introduced into the lattice. Moreover, limitations on topological charge of stationary vortex-carrying states that exist in structures with discrete rotational symmetry~\cite{49} hint that some linear vortices observed in~\cite{52} are essentially dynamical states, whose reshaping and modification of phase structure is not seen due to short propagation distances.

In this article, we report on the first experimental observation of topological vortex solitons in disclination arrays with $\mathcal{C}_5$ discrete rotational symmetry obtained by introducing a crystallographic defect (removing a sector) into original $\mathcal{C}_6$ honeycomb array. The disclination arrays with well-defined disclination core in topological and non-topological phases were inscribed in fused silica using fs-laser direct writing technology adjusted to create waveguides supporting circularly-symmetric modes. We observed drastic differences in localization dynamics in non-topological arrays, in which the formation of vortex solitons with topological charge $m=1$ occurs only in the semi-infinite gap above a considerable power threshold due to the absence of topological in-gap modes, and in topological arrays, in which $m=1$ vortex solitons bifurcating from localized topological states are thresholdless and show stability at any power level in agreement with theoretical predictions.

\begin{figure*}[t]
\centering
\includegraphics[width=1.00\linewidth]{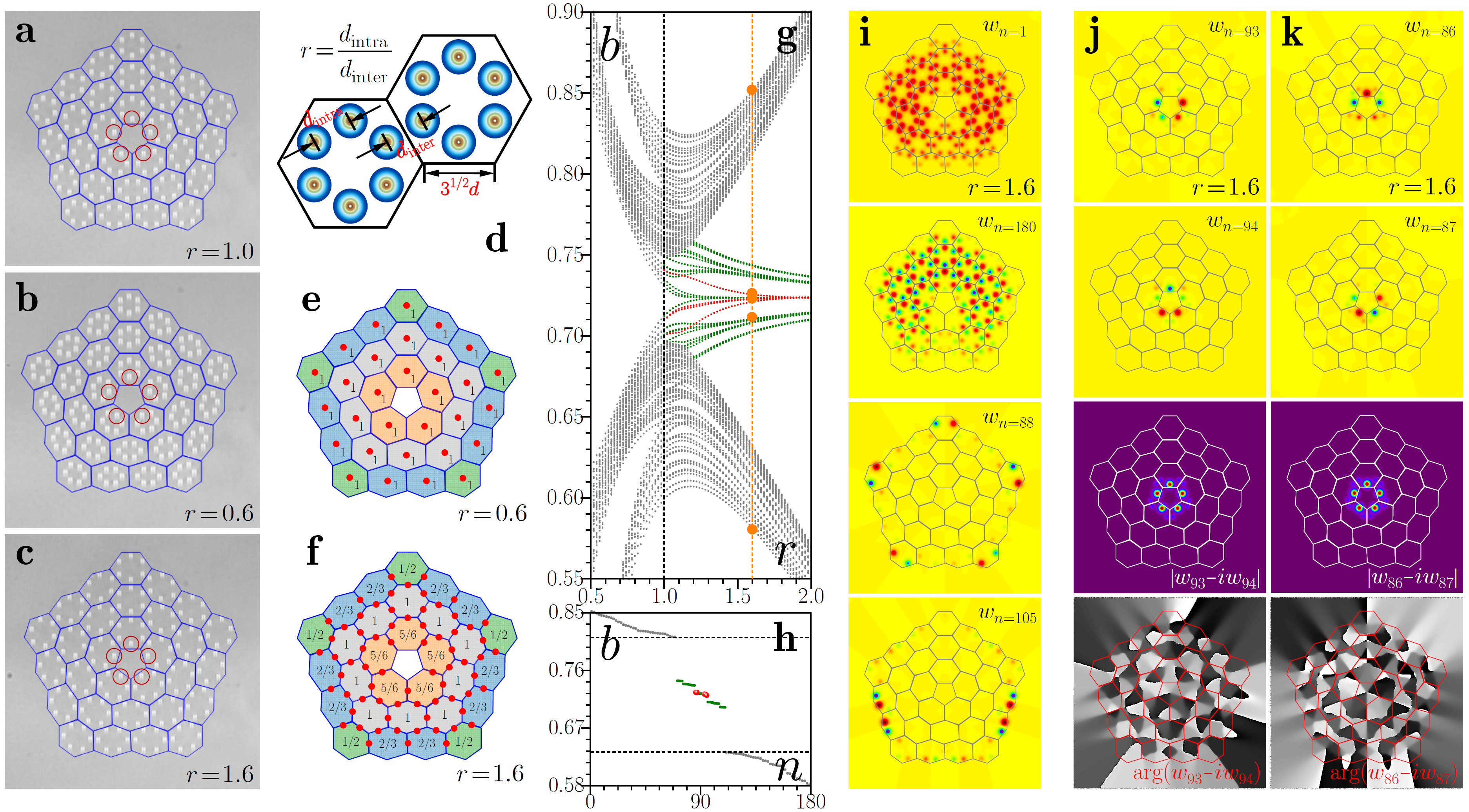}
\caption{{\bf Topological modes in disclination arrays.} ({\bf a-c}) Microphotographs of disclination arrays with $\mathcal{C}_5$ discrete rotational symmetry and different distortion parameters $r$. The red circles mark the waveguides on the disclination core, which were excited in the experiments. Blue lines highlight unit cells of the structure. ({\bf d}) Schematic illustration of two cells of the original honeycomb array used to construct the disclination array, with indication of intracell and intercell spacings. Positions of the Wannier centers (red dots) and spectral charges for all cells in the trivial array with $r=0.6$ ({\bf e}) and in the topological array with $r=1.6$ ({\bf f}). ({\bf g}) Linear spectrum of the disclination array vs distortion parameter $r$. Gray dots -- bulk modes, green dots – outer edge and corner states, red dots -- disclination states. The large orange dots correspond to the states shown in ({\bf i-k}). ({\bf h}) Eigenvalues $b_n$ of the modes vs mode index $n$ at $r=1.6$ corresponding to the orange dashed line in ({\bf g}). ({\bf i}) Examples of bulk states, and outer corner and edge states $w_n$ at $r=1.6$. Panels ({\bf j}) and ({\bf k}) illustrate the profiles of the degenerate disclination states $w_n$, whose linear combinations generate vortex disclination states with topological charges $m=1$ and $m=2$, whose modulus and phase distributions are shown in the two lower panels. Blue lines in ({\bf a})-({\bf c}) and gray lines in ({\bf i})-({\bf k}) indicates array cells. Here and in all following figures $d=3.2$, $p=3.1$, $a_x=0.72$, $a_y=0.78$.}\label{fig1}
\end{figure*}

\section*{Results}\label{sec1}

HOTIs with topological disclinations can be constructed from periodic structures, e.g. finite honeycomb waveguide arrays originally possessing $\mathcal{C}_6$ discrete rotational symmetry. The construction method illustrated in Fig.~1(a)-1(d) involves a two-step procedure. At the first step, the Kekul\'{e} distortion is introduced into all unit cells of the honeycomb array with the original spacing between the waveguides $d$ by shifting each waveguide in the direction perpendicular to the nearest side of the unit cell. Such a distortion is characterized by the parameter $r=d_{\text{intra}}/d_{\text{inter}}$, where $d_{\text{intra}}$ and $d_{\text{inter}}$ are the final intra- and intercell waveguide spacings [Fig.~1(d)]. The shift affects the coupling strengths between the waveguides, making intercell couplings stronger than intracell ones at $r>1$ and weaker than intracell ones at $r<1$. At the second step, one can remove (or add) the angular sector with Frank angle $2\pi \ell/6$ ($\ell=1,2,\ldots$) from the array and adjust the positions of the waveguides in the remaining cells so that they fill/accommodate the removed/added sector. This procedure produces a disclination array that has a discrete rotational symmetry $\mathcal{C}_{6\pm \ell}$ that may not be accessible in periodic structures (this is the case, for example, for $\ell=1$ that corresponds to $\mathcal{C}_5$ or $\mathcal{C}_7$ aperiodic arrays). Figures~1(a)-1(c) show microphotographs of $\mathcal{C}_5$ waveguide arrays with disclination and different $r$ values designed using this method and inscribed in $10$~$\textrm{cm}$ slabs of fused silica (the examples of disclination arrays with other symmetries can be found in \cite{43,44,48}). Here we use multiscan writing technology (see {\bf Methods} for details of array fabrication), where each waveguide is created by six closely written tracks. Such waveguides remain single-mode, while their eigenmodes are practically circularly symmetric, which eliminates the anisotropy of the coupling (isotropic couplings are desirable for degeneracy of the topological states at the disclination core used for vortex construction below). As one can see, this procedure creates a disclination core in the center of the array with five waveguides (red circles) lying on the core. We have inscribed sufficiently large arrays with $N=180$ waveguides to ensure that there is no coupling between the states at the disclination core and topological modes that may form in the outer corners.

Non-trivial topological properties in disclination arrays emerge upon increase of the distortion parameter $r$ and are associated with a fractional spectral charge, which in the language of solid-state physics is an indication of the filling anomaly \cite{42,43,44,46,47}. The spectral charge $\mathcal{Q}$ can be evaluated by integrating the local density of all states that lie below the bandgap---this allows to make a conclusion about the number of modes (or populated sites) per unit cell. A more straightforward approach to evaluation of spectral charge per unit cell utilizes the analysis of the Wannier centers positions (which would indicate the average positions of the electrons in electronic systems with a similar structure). For $r<1$ they are located in topologically trivial positions in centers of the cells [see red dots in Fig.~1(e)] yielding unit spectral charge per unit cell. By contrast, at $r>1$ the Wannier centers are located in the middle of the edges between cells (i.e., they are shared by neighboring cells), except for the outer edges of the cells at the boundary of the array [red dots in Fig.~1(f)], because charge fractionalization is a property of the bulk. In this case, count of fractionalized Wannier centers and normalization to the number of occupied sites in the bulk, yields a unit charge for the bulk cells and a fractional charge of $5/6$ for all cells at the disclination core. This filling anomaly indicates that the disclination core can support localized modes of topological origin at $r>1$ (which coexist with the corner and edge states at the outer boundary of the array, because the latter cells are also characterized by the fractional charges $1/2$ and $2/3$, respectively). A qualitative change in the linear spectrum of the array modes is therefore expected at the transition from the trivial ($r<1$) to topological ($r>1$) regime.

\section*{Linear spectrum and vortex solitons at the disclination core}\label{sec2}

The propagation of paraxial light beams in our shallow waveguide arrays is described by the nonlinear Schr\"{o}dinger equation for the dimensionless amplitude of the light field $\psi$:
\begin{align}  
\label{eqNLS}
    i\frac{\partial\psi}{\partial z} = -\frac{1}{2}\left(\frac{\partial^2 \psi}{\partial x^2} + \frac{\partial^2 \psi}{\partial y^2}\right) -\mathcal{R}(x,y)\psi - |\psi|^2\psi.
\end{align}
where $x$, $y$, and $z$ are the dimensionless transverse coordinates and propagation distance, respectively, the waveguide array with disclination is described by the function $\mathcal{R}(x,y)=p\sum_{k,l}\mathcal{V}\left(x-x_k,y-y_l\right)$ and consists of Gaussian waveguides $\mathcal{V}=e^{-x^2/a_x^2-y^2/a_y^2}$ located in the nodes with coordinates $x_k$, $y_l$. Here we account for a slight residual ellipticity of the waveguides $a_x=0.72$ and $a_y=0.78$ (corresponding to $7.2\times7.8$~$\mu\textrm{m}^2$ waveguides) for the best agreement with the experiments, set the original spacing between waveguides $d=3.2$ ($32$~$\mu\textrm{m}$) and the array depth $p=3.1$, which is defined by the refractive index contrast in the laser-written structure. All normalizations are presented in {\bf Methods}.

Omitting cubic nonlinearity in (1), we obtain linear eigenmodes as $\psi=w e^{ibz}$, where $w$ is the real function describing the modal shape and $b$ is the eigenvalue. The eigenvalues of all modes of the array, which are sorted such that $b=b_n$ decreases with increasing mode index $n$, are shown in Fig.~1(g) vs distortion parameter $r$. In the trivial phase ($r<1$) there are no states in the gap between bulk bands (gray dots) and all modes of the array are extended. In agreement with above analysis of spectral charges, in the topological phase ($r>1$) localized modes appear in the gap between the bulk bands: The green lines correspond to corner and edge states at the outer edge of the array [see examples of profiles in Fig.~1(i)], while the red lines correspond to modes localized at the disclination core [see profiles in Fig.~1(j) and~1(k)]. There are five disclination modes in the spectrum in topological phase, as dictated by $\mathcal{C}_5$ discrete rotational symmetry of the array, see red dots in Fig.~1(h) showing eigenvalues at $r=1.6$. The spectrum shows the presence of two degenerate pairs of disclination modes, whose linear combinations, $w_{93}\mp iw_{94}$ and $w_{86}\mp iw_{87}$, can produce compact vortex-carrying modes with topological charges $m=\pm1$ [Fig.~1(j)] and $m=\pm2$ [Fig.~1(k)]. Notice that small residual ellipticity of the fabricated waveguides that we take into account here to achieve best agreement with experimental results, slightly lifts the degeneracy of disclination states in comparison with the ideal case of circular waveguides with $a_x=a_y$. For our waveguide parameters $p=3.1$, $a_x=0.72$, $a_y=0.78$, the difference between eigenvalues of modes giving rise to $m=\pm 1$ vortex state amounts to $b_{93}-b_{94}\approx 3.2\times 10^{-4}$, while this difference for modes producing $m=\pm 2$ vortex amounts to $b_{86}-b_{87}\approx 9.8\times 10^{-6}$. This is a negligible difference, since it leads to beating lengths exceeding sample length by two orders of magnitude, and can be compensated even by a weak nonlinearity.

The structure of the linear spectrum thus reflects the fundamental constraint imposed by the discrete rotational symmetry of the array on the charge of the disclination vortex, which can exist in the array in \textit{stationary form}, see \cite{53}. Therefore, arrays with lower order of discrete rotational symmetry $\mathcal{C}_{3,4}$ can only support \textit{stationary} disclination vortices with $|m|=1$, while $\mathcal{C}_{7,8}$ arrays can support vortices with charges up to $|m|=3$ (see {\bf Supplemental Materials} for example of linear spectrum of the $\mathcal{C}_7$ array, its modes and possible linear combinations producing vortex states on disclinations, as well as for discussion of properties of topological vortex solitons in such systems with higher discrete rotational symmetry). Note that due to the staggered tails reflecting the topological nature of the vortex states, they have very complex phase distributions, as can be seen in the bottom panels in Fig.~1(j) and~1(k), where in addition to the central phase singularity, one can observe several off-center singularities within the cells surrounding the disclination core. The localization of such topological modes on the disclination core increases with the increase of the distortion parameter $r$.

\begin{figure*}[t]
\centering
\includegraphics[width=1.0\textwidth]{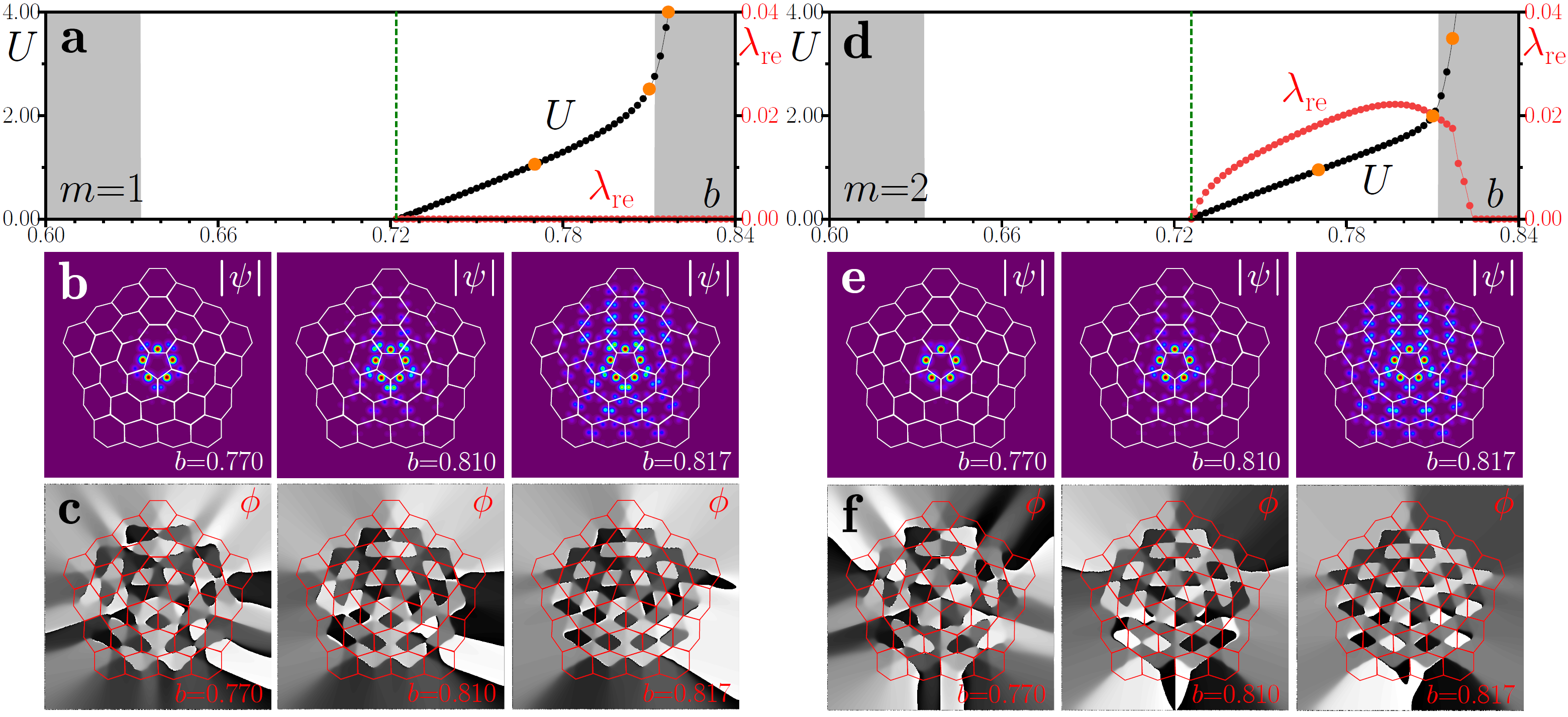}
\caption{{\bf Families of topological vortex solitons on disclination.} Dependencies of the power $U$ on the propagation constant $b$ (black dots) and examples of soliton field modulus $|\psi|$ and phase $\phi=\arg\left(\psi\right)$ distributions for different propagation constant values for solitons with topological charge $m=1$ ({\bf a-c}) and $m=2$ ({\bf d-f}). In ({\bf a}), ({\bf d}) gray regions indicate bands, vertical dashed lines indicate the eigenvalue of the linear disclination states from which the vortex soliton family bifurcates. Soliton profiles correspond to the orange dots in ({\bf a}), ({\bf d}). The red dots in ({\bf a}),({\bf d}) show the dependencies of the perturbation growth rate $\lambda_{\text{re}}$ on $b$ illustrating stability of the $m=1$ vortex solitons and instability of the $m=2$ solitons. White and red grid in ({\bf b}),({\bf c}) and ({\bf e}),({\bf f}) indicate array cells.}\label{fig2}
\end{figure*}

Topological vortex solitons bifurcate from the linear vortex states constructed above due to focusing nonlinearity. They were found from Eq.~(1) in the form $\psi=w e^{ibz}$ using Newton method, where $w$ is the complex function and $b$ is the propagation constant defining the soliton power $U=\iint|w|^2dxdy$. The families $U(b)$ of topological vortex solitons with $m=1,2$ are shown in Fig.~2(a) and~2(d), respectively. To illustrate the feasibility of the experimental observation of vortex solitons, we have performed a linear stability analysis for them (see {\bf Methods}). In Fig.~2(a),(d), alongside the soliton families, we also plot the growth rate $\lambda_{\text{re}}$ (red dots) for the most destructive perturbation. The vortex soliton is stable if $\lambda_{\text{re}}\leq 0$, but $\lambda_{\text{re}}> 0$ means instability, since at least one perturbation can grow exponentially (i.e., the amplitude of such perturbation will increase $\sim e^{\lambda_\textrm{re}z}$ and eventually it will cause considerable deformation of the vortex soliton). The bifurcation points (different for $m=1$ and $m=2$), where the power of the vortex soliton vanishes are indicated by dashed green lines in Fig.~2(a),(d). Due to their topological nature, such solitons are thresholdless, in contrast to vortex solitons in non-topological arrays. With increasing power the propagation constant shifts towards the upper edge of the gap and eventually into the band (gray region). This is accompanied by a broadening of the vortex solitons and their eventual delocalization [see $|\psi|$ distributions in Figs.~2(b),(e)], while the vorticity is always seen in phase distributions $\phi=\arg\left(\psi\right)$ [Fig.~2(c),(f)]. Remarkably, the stability analysis predicts compete stability of the $m=1$ family, even inside the band, and an instability of the $m=2$ family [see $\lambda_{\text{re}}(b)$ dependencies]. Similar conclusion was obtained also for vortex solitons in $\mathcal{C}_7$ disclination arrays, where only solitons with $m=1$ were found to be stable, see {\bf Supplemental Materials}. This is also in contrast to the stability properties of vortex solitons in non-topological arrays with focusing nonlinearity, where only high-charge vortices are usually stable, while low-charge vortices are unstable \cite{40}. We also studied robustness of vortex solitons in the presence of disorder introduced into disclination structure (into depths and positions of individual waveguides). Such robustness was studied by propagating vortex solitons in perturbed arrays over very large distances, exceeding sample length by several orders of magnitude. We found that $m=1$ solitons survive in the presence of such disorder at any power level, while in the case of $m=2$ solitons the disorder usually leads to instabilities. It should be emphasized that the vorticity-carrying nonlinear states described here are fundamentally different from localized linear modes with trivial phases that form on vortex-like distortions introduced directly into the array structure \cite{53}.

\begin{figure*}[t]
\centering
\includegraphics[width=1.0\textwidth]{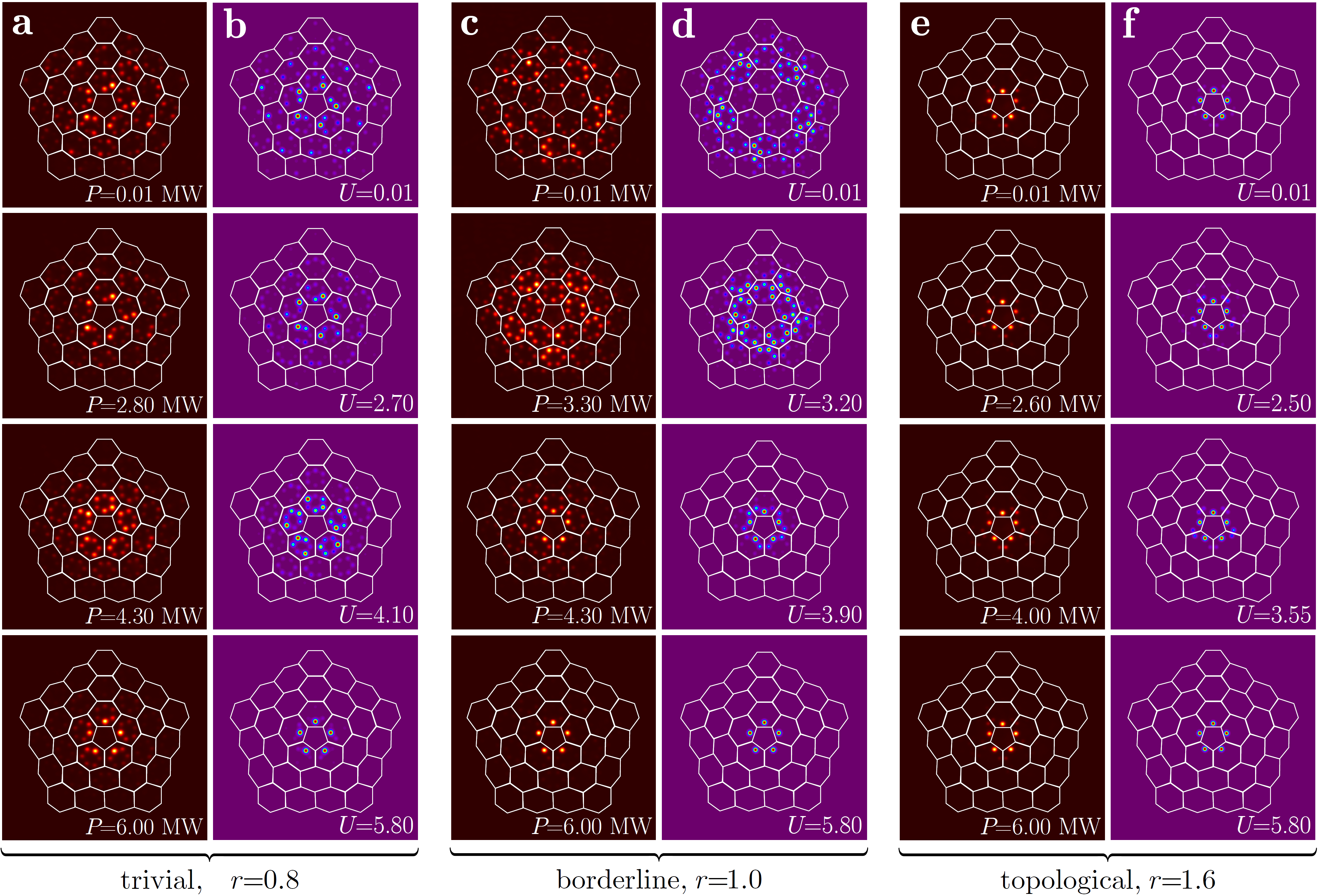}
\caption{{\bf Excitation of vortex solitons on the disclination core.} Experimentally measured [columns ({\bf a}),({\bf c}),({\bf e}) with marron background] and corresponding numerically calculated [columns ({\bf b}),({\bf d}),({\bf f}) with violet background] intensity distributions at the output face of $10$~$\textrm{cm}$ sample illustrating dynamical excitation of vortex solitons with charge $m=+1$ above considerable power threshold in arrays in trivial ({\bf a}),({\bf b}) (corresponding to $r=0.8$) and borderline ({\bf c}),({\bf d}) (corresponding to $r=1.0$) phases, as well as the formation of thresholdless vortex solitons at the disclination core of array in topological phase ({\bf e}),({\bf f}) (corresponding to $r=1.6$). Input power levels are indicated at each experimental and theoretical panel. White lines indicate array cells. Phase distributions corresponding to theoretical intensity distributions are shown in Extended Data Fig.~1.}\label{fig3}
\end{figure*}

\begin{figure*}[t]
\centering
\includegraphics[width=0.75\textwidth]{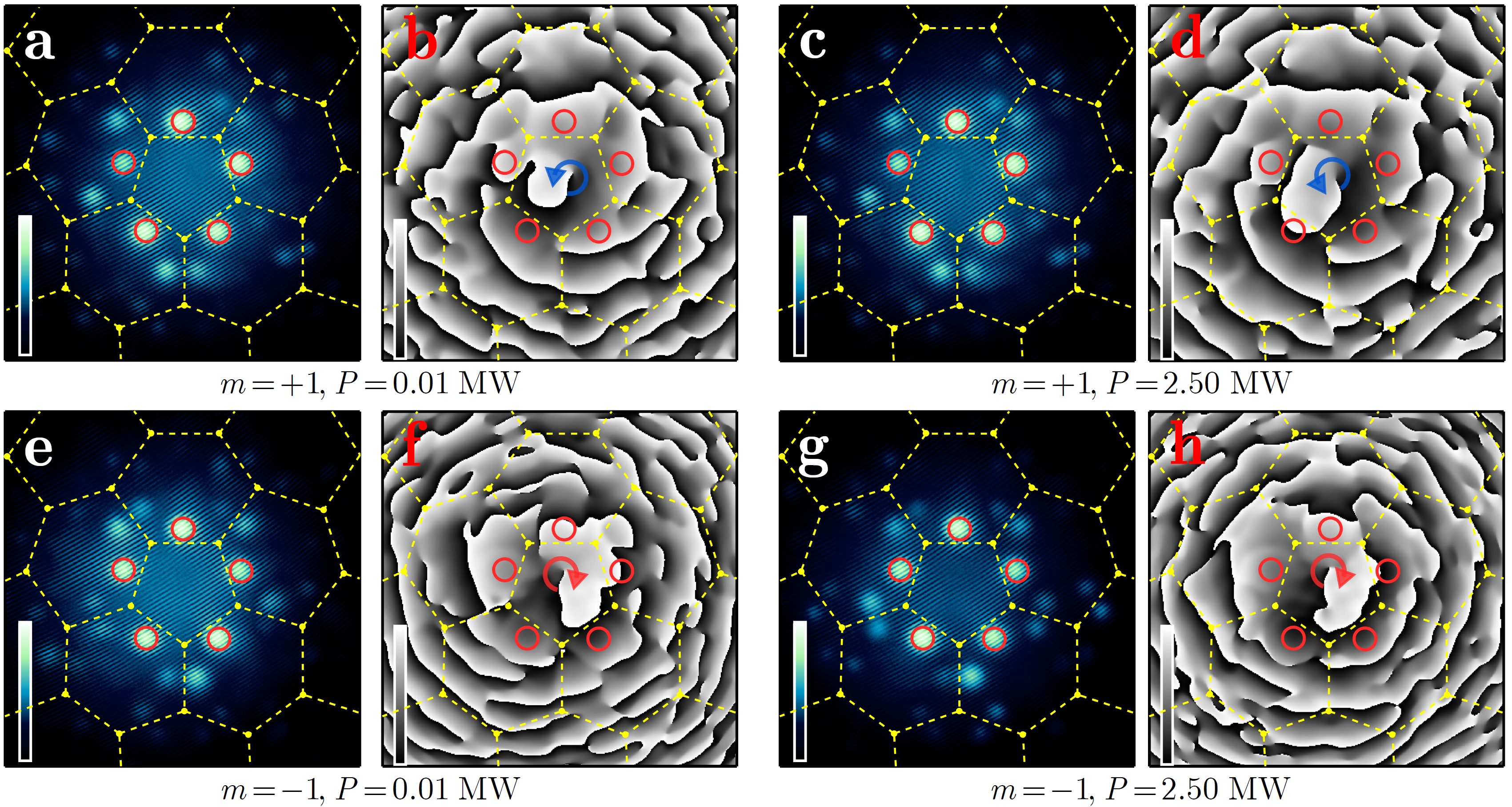}
\caption{{\bf Experimental proof of the formation of topological vortex soliton at the disclination core at  $r=1.6$.} Single-shot interference patterns with normalized intensity distribution ({\bf a}),({\bf c}) and corresponding retrieved phase distributions ({\bf b}),({\bf d}), confirming the formation of vortex states with topological charge $m=+1$ (blue arrow) at low power $P=0.01$~$\text{MW}$ ({\bf a}),({\bf b}) and at sufficiently high power $P=2.50$~$\text{MW}$ ({\bf c}),({\bf d}). When the array is illuminated with five spots with the opposite phase winding, the state with opposite topological charge $m=-1$ (red arrow) is excited, as can be seen from ({\bf e}),({\bf f}) at $P=0.01$~$\text{MW}$ and ({\bf g}),({\bf h}) at $P=2.50$~$\text{MW}$. The color scale in panels ({\bf a}),({\bf c}),({\bf e}), and ({\bf g}) is logarithmic and saturated at the $0.08$ level to visualize the fork-like dislocation between five excited waveguides (marked with red circles). Yellow dashed lines schematically show the array cells.}\label{fig4}
\end{figure*}

\section*{Observation of topological vortex solitons}\label{sec3}

To observe topological vortex solitons on disclinations in our fs-laser written waveguide arrays [Fig.~1(a)-1(c)], we used $1.5$~$\text{ps}$ pulses of variable energy $E$ derived from a $1$~$\text{kHz}$ Ti:Sa laser system at the central wavelength of $800$~$\text{nm}$. This allows to achieve peak powers $P$ up to several $\text{MW}$, at which the focusing nonlinearity of fused silica strongly influences the propagation of light. To excite vortex states, we used a reflective spatial light modulator (SLM), which transforms incident laser radiation to a coherent light field distribution consisting of five spots with adjustable width, phase and intensity positioned at the vertices of a regular polygon (equally spaced on a ring of the desired radius). This structured laser light is focused on the sample to excite five waveguides at the disclination core, which are indicated with red circles in Fig.~1(a)-1(c). To add a global vorticity to the input field, the controllable phase of $2\pi m(\ell-1)/5$ was imposed on each spot (here spot index $\ell=1,\ldots,5$ increases in the counterclockwise direction). This leads to a total accumulation of the phase of $2\pi m$ on the closed contour surrounding five spots, resulting in efficient excitation of the vortex states due to the strong overlap of such an input with modal combinations such as $w_{93}\mp iw_{94}$ (generating $m=\pm1$ vortex) or $w_{86}\mp iw_{87}$  (generating $m=\pm2$ vortex) (see {\bf Methods} for the details on the experimental excitation and detection of vortex solitons).

In Fig.~3 we compare experimentally observed output intensity distributions (images with maroon background) with results of theoretical simulations (images with violet background) for five-spot inputs with different powers and $m=+1$ in arrays with different distortion parameters $r$. In the theoretical images, the input power is defined as $U=\iint|\psi|^2dxdy$. In the array with $r=0.8$ (in the trivial phase), such input at low and moderate powers results in strong diffraction of light over the entire structure due to the absence of localized modes in this structure, as can be seen in Fig.~3(a),(b). The presence of vorticity in the light beam is evidenced by a specific ``angular twist'' of the output pattern, which is clearly visible at $P=0.01$~$\text{MW}$ and $P=2.8$~$\text{MW}$ that changes to the opposite one when the charge $m$ changes sign. The pattern nevertheless features $\mathcal{C}_5$ discrete rotational symmetry of the array, indicating that the modal fields of individual waveguides are practically circular and the coupling is isotropic. The intensity distribution gradually contracts to the disclination core as the input power increases, but even at $P\sim6$~$\text{MW}$ the localization is still moderate, indicating that in a trivial array vortex solitons at the disclination form only above a considerable power threshold and with propagation constants belonging to the semi-infinite gap laying above all gray bands in Fig.~1(g) (as it occurs also in common periodic lattices). In the borderline case $r=1.0$, presented in Fig.~3(c),(d), similar dynamics are observed. At low and moderate powers, even a stronger expansion of the light beam across the array is visible, but already at powers $P\sim4$~$\text{MW}$ the light contracts to the disclination core, indicating the formation of a vortex soliton, also from a semi-infinite gap. For this $r$ value, the power threshold for soliton formation is therefore slightly reduced.

A qualitatively different dynamics is observed for a topological array with $r=1.6$, as shown in Fig.~3(e),(f). In this case, our input excites strongly localized topological vortex solitons at the disclination core in a wide range of input powers, starting from minimal power levels, i.e., such solitons are clearly thresholdless in agreement with theoretical predictions of Fig.~2. The theoretical output phase distributions shown in Extended Data Fig.~1, which correspond to the intensity distributions in Fig.~3, clearly confirm the formation of the $m=+1$ vortex soliton with phase singularity at the center of the disclination core. The global vorticity nested in the field is clearly visible in these phase distributions (it is present even in non-topological arrays when strong diffraction occurs). The phase distributions at $r=1.6$ obtained at low and moderate powers [panel (c) of Extended Data Fig.~1, $U=0.01$ and $U=2.5$] contain several phase singularities surrounding a central one that is representative, namely for topological vortex solitons from the forbidden topological gap [c.f. Fig.~2(c)]. However, this phase structure transforms into a different structure with the single singularity at the center with increase of power [panel (c) of Extended Data Fig.~1, $U=3.55$]. We attribute this to the fact that, according to Fig.~2(a), at $U\approx2.8$ a topological soliton enters the band, where the coupling with bulk modes takes place. This is accompanied by a slight expansion (due to the finite sample length, only the initial stage of this process can be seen) of the output intensity distribution visible, for example, in Fig.~3(e) at $P=4$~$\text{MW}$. Further increase of power to $6$~$\text{MW}$ results in the excitation of a vortex soliton in a semi-infinite spectral gap, similar to solitons observed at high powers in the non-topological array with $r=1.0$. Thus, our setup enables the observation of the transition between the excitation of vortex solitons from different spectral gaps, with different phase structure.

These differences in excitation dynamics in topological and non-topological arrays find their manifestation in qualitatively different dependencies of output form-factor $\chi=[U^{-2}\iint|\psi|^4dxdy]^{1/2}$ (this quantity, inversely proportional to beam width, characterizes localization of the pattern with better localization corresponding to higher values of $\chi$) on input power, as shown in Fig.~S1 of {\bf Supplemental Materials}.

At the same time, we note that $m=\pm2$ vortex solitons (that are also allowed by the discrete rotational symmetry of our array) were not observed experimentally due to their intrinsic instability even in the topological array described in Fig.~2. Under the action of perturbations unavoidable in experiment, we always observe splitting of charge-$2$ phase singularity in the center into charge-$1$ singularities and the onset of azimuthal modulation instability for such states, leading to irregular oscillations of the bright spots.

To unambiguously prove the formation of $m=+1$ topological vortex solitons, we have performed single-shot interferometric measurements. Figures~4(a),(c) show typical examples of the observed interference patterns between the output vortex field and the resonant reference plane wave (at zero-time delay between the arms of the Mach-Zehnder interferometer) for different input powers. For more details, see {\bf Methods}. Figure~4(a) corresponds to the low power $P=0.01$~$\text{MW}$, at which quasi-linear localized vortex state forms, while Fig.~4(c) corresponds to the high power $P=2.5$~$\text{MW}$, at which vortex soliton with propagation constant located close to the upper border of the topological gap is expected to form. The interference patterns (intentionally plotted in logarithmic color scale) reveal fork-like phase singularities as well as a representative structure of topological vortex -- five most intense spots on the disclination core are surrounded by pairs of less intense spots belonging to different cells. Extracted corresponding phase distributions in Fig.~4(b),(d) confirm the formation of the vortex within disclination core, as well as additional singularities near the borders of the surrounding cells - a signature of topological vortex states. When the sample is illuminated by five spots excitation pattern with opposite phase winding, we observe the formation of $m=-1$ topological vortex solitons, as expected. Single-shot interference patterns and the corresponding extracted phase distributions are shown in Fig.~4(e)-(h) for the same power levels, confirming the equivalence of the properties of states with opposite charges $m$.

\section*{Discussion}\label{sec4}

We have provided the first experimental evidence of topological soliton carrying vorticity excited in aperiodic waveguide arrays with an inner disclination core. Discrete rotational symmetry of such structures is directly manifested in the structure of the linear spectrum and the number of topological modes that can be guided by the system. The observed solitons are characterized by several distinct features. First, they are thresholdless, and second, their phase structure and stability properties differ dramatically from those of usual vortex lattice solitons. Based on the presented results, we envision a non-trivial interplay between discrete rotational symmetry and topology for nonlinear processes occurring in disclination structures. For example, parametric interactions (especially when they involve vortex-carrying states) are expected to be strongly influenced by this interplay (apart from general enhancement of interactions due to localization at the disclination core). The interaction between topology and nonlinearity can be particularly intriguing in driven-dissipative systems (such as polariton condensates in structured microcavities), where vorticity can be imposed on the condensate by the external pump or spontaneously formed as a product of mode competition, resulting in non-trivial dynamics. The strong confinement of topological vortex states on the disclination core in the system affording topological protection can be useful for the realization of various switches and quantum memory. We believe that our results can also be employed for improving the stability and design of topological vortex lasers, as well as for future applications connected with information transmission in compact topological states.


\section*{Methods}\label{sec5}

\subsection*{Fs-laser inscription of disclination arrays}\label{sub1}

Waveguide arrays with disclinations were written in ${10\,\textrm{cm}}$ long fused silica substrate (JGS1). Circularly polarized laser beam with a central wavelength of $515\,\textrm{nm}$, pulse duration of ${230\,\textrm{fs}}$, repetition rate of ${1\,\textrm{MHz}}$, and pulse energy of ${270\,\textrm{nJ}}$ was focused by an aspheric lens (${\textrm{NA} = 0.4}$) into the bulk of the substrate. We have verified that with such relatively soft focusing the spherical aberrations contribute negligibly when waveguides are fabricated around the preselected optimal depth (see details in {\bf Supplemental Materials}), enabling the fabrication of homogeneous waveguide arrays within the depth range from $400$ to $1000\,\mu\textrm{m}$. The substrate was translated relative to the laser beam using a high-precision positioning system (AeroTech).

For a given $\rm NA$ and a typical scanning velocity of $1~\textrm{mm/s}$, the individual waveguides inscribed with these parameters would have a significantly elongated cross-section and an elliptical eigenmode with aspect ratio around $0.7$. {This is because the modification of the material occurs within the focal region of the writing laser beam that is elongated in the direction of the writing beam optical axis. Since waveguides are written in the direction orthogonal to the writing beam's axis, their cross-sections would be also elliptical with an aspect ratio depending on the $\rm NA$ of the focusing lens. The coupling between such waveguides would be anisotropic, which would lead to a reduction in the symmetry of the entire system. Various approaches to the creation of waveguides with circular cross-sections were suggested that utilize beam shaping techniques based on mechanical slit, cylindrical optics, or spatial light modulator~\cite{56,57,58}. However, in such cases the size of the focal area usually increases significantly, which necessitates an increase in writing energy and a reduction in writing speed, thereby increasing writing time and making these approaches impractical for large waveguide arrays, such as arrays considered here.}

{To overcome this problem, here we have applied a multiscan approach~\cite{59} that allows for faster writing of the waveguide array and also considerably reduces the sensitivity to fabrication errors. Using this approach we have designed waveguides with rectangular cross-section and nearly-circular horizontally polarized eigenmode with aspect ratio of $0.98$.} Each waveguide in the array was composed of 6 mutually displaced tracks with a displacement of $1.6\,\mu\textrm{m}$. The tracks were ordered during writing from the center of the waveguide toward the edge, with the writing direction being altered for each track. Tracks were written at $30\,\textrm{mm/s}$, resulting in an effective writing velocity of $5\,\textrm{mm/s}$ and reducing writing time for a single structure to 1.2 hours. This approach drastically reduces the anisotropy of the coupling and improves the symmetry of the observed vortex modes on disclination. Rectangular waveguides exhibit a propagation loss of ${0.1\,\textrm{dB/cm}}$ at ${\lambda = 800\,\textrm{nm}}$.

\subsection*{Experimental excitation and detection of vortex solitons}\label{sub2}

To excite vortex solitons, we used a femtosecond Ti:Sa laser system (Spitfire Pro, Spectra Physics) operating at 800 nm with a pulse repetition rate of 1 kHz. To reduce the influence of effects associated with phase self-modulation in the nonlinear regime and to minimize the dependence of the evanescent coupling strength on the wavelength, initially short 40 fs pulses were spectrally narrowed down to 10 nm using an interference filter (FB800-10, Thorlabs) and stretched to 1.5 ps (negatively chirped) using built-in grating compressor. After passing through the beam steering system (Avesta), the laser radiation was transformed using a reflective phase-only spatial light modulator (SLM). The SLM (PLUTO-2.1-NIR-118 Holoeye) operating in the first order of diffraction allows us to realize simultaneous excitation of five waveguides in controllable manner. Namely, using  programmability of the phase mask, we control a distance between the excitation spots (at the vertices of a regular polygon), their size, individual intensity and relative phase - a crucial parameter for the effective excitation of the vortex soliton. The multi-spot excitation pattern was spatially filtered out with a vertical slit (to cut-off other parasitic diffraction orders) in the 4f lens configuration. The spatially transformed laser excitation was focused (with an aspherical lens with a focal length of 100 mm) onto a sample mounted on a 6-axis nanopositioner (I6000 6-Axis XYZ/RYP, Luminos). The calculated overlap integral for this lens and the waveguide mode is 0.85, while the positioning accuracy was better than 1 $\mu$m.

In order to precisely control the relative intensities and the phases in the excitation channel, we utilized a modified Mach-Zehnder interferometer. For this, using a beam sampler we reflect a small fraction of the excitation laser beam before it hits the SLM and then expand it to create a reference plane wave. This plane wave is then interfered with the structured excitation profile at zero time delay between the interferometer arms. The product of the single-shot interference is detected with a fast CMOS camera and then analyzed using the off-axis digital holography technique~\cite{60,61}.

To measure the intensity distribution and the phase map distribution at the output of the sample with inscribed waveguides, we use 
additional scientific CMOS camera (Kiralux 12.3 Mp, Thorlabs) and the second Mach-Zehnder interferometer. For this, we expand ($\geq$20$\times$) the beam corresponding to one of the waveguides (at the rear facet), which serves as a reference plane wave, and interferes with the array output intensity distribution. The typical examples of the single-shot interference patterns and the retrieved phase distributions are given in Fig.~4 in the main text.

\subsection*{Normalization of parameters in theory}\label{sub3}

The theoretical description of the formation of vortex solitons is based on the dimensionless nonlinear Schr\"{o}dinger equation~(1). In this equation, the dimensionless transverse coordinates $x=X/r_0$, $y=Y/r_0$ are normalized to the characteristic transverse scale $r_0=10$~$\mu\text{m}$, the dimensionless propagation distance $z=Z/L_d$ is normalized to the diffraction length $L_d=kr_0^2\approx1.14$~$\text{mm}$, where $k=2\pi n/\lambda$ is the wavenumber at the working wavelength $\lambda=800$~$\text{nm}$, $n\approx1.45$ is the unperturbed refractive index of fused silica. The dimensionless field amplitude is related to the real field amplitude $\mathcal{E}$ via $\psi=\left(k^2r_0^2n_2/n\right)^{1/2}\mathcal{E}$, where the nonlinear refractive index $n_2\approx2.7\times10^{-20}$~$\text{m}^2/\text{W}$. The dimensionless depth of the optical disclination potential $\mathcal{R}$ is given by $p=k^2r_0^2\delta n/n$, where $\delta n$ is the refractive index modulation depth. The value $p=3.1$ used in the main text therefore corresponds to a refractive index modulation $\delta n\approx3.5\times10^{-4}$. The width of the waveguides $a_x=0.72$, $a_y=0.78$ corresponds to $7.2$~$\mu\textrm{m}$ and $7.8$~$\mu\textrm{m}$, respectively, while the spacing $d=3.2$ in the original honeycomb array used for construction of the disclination structure corresponds to $32$~$\mu\textrm{m}$. The length of sample of $10$~$\text{cm}$ corresponds to the dimensionless propagation distance $z\approx88$.

\subsection*{Stability analysis for vortex solitons}\label{sub4}

To prove that vortex solitons can be stable, we performed linear stability analysis for the obtained solutions by substituting perturbed vortex state $\psi=\left[w(x,y)+u(x,y)e^{\lambda z}+v^*(x,y)e^{\lambda^*z}\right]e^{ibz}$, where $u$,$v\ll w$ are small perturbations and $\lambda=\lambda_{\text{re}}+i\lambda_{\text{im}}$ is the complex perturbation growth rate, into Eq.~(1). Its linearization around the stationary vortex soliton solution $w$ results in the linear eigenvalue problem
\begin{equation} \label{stability}
    \begin{split}
    \lambda u&=+i\left[(1/2)\Delta u+\mathcal{R}u-bu+2|w|^2u+w^2v\right], \\
    \lambda v&=-i\left[(1/2)\Delta v+\mathcal{R}v-bv+2|w|^2v+w^2u\right],
    \end{split}
\end{equation}
that was solved numerically. In Eq.~(2) $\Delta=\partial^2/\partial x^2+\partial^2/\partial y^2$. Vortex soliton $w$ is stable when $\lambda_{\text{re}}\leq0$ for all possible perturbations and is unstable when $\lambda_{\text{re}}>0$ at least for one of them.



\subsection*{Data availability}

Data supporting conclusions of this work are included within the Article and its Supplementary Information. All other raw data that support the findings of this study are available from the corresponding author on reasonable request.

\subsection*{Acknowledgements}

We are grateful to Philipp Grigoriev for useful discussions.
This work was supported by the Russian Science Foundation (grant 24-12-00167) and partially by the research project FFUU-2024-0003 of the Institute of Spectroscopy of the Russian Academy of Sciences. S.A.Z. acknowledges support by the Foundation for the Advancement of Theoretical Physics and Mathematics ``BASIS'' (22-2-2-26-1).
S.K.I. has received funding from the European Union through the Program Fondo Social Europeo Plus 2021-2027(FSE+) of the Valencian Community (Generalitat Valenciana CIAPOS/2023/329).
This project was developed within the framework of the General
Collaboration Agreement between the University of Valencia
and Banco Santander, S.A., for the Santander Postdoctoral Research Scholarships (las becas Santander Investigaci\'{o}n
Postdoctoral).

\subsection*{Authors contributions}

A.V.K. and K.S. contributed equally to this work. All authors contributed significantly to this work.

\subsection*{Competing interests}

The authors declare no conflicts of interest.

\begin{figure*}[t]
\centering
\includegraphics[width=0.7\textwidth]{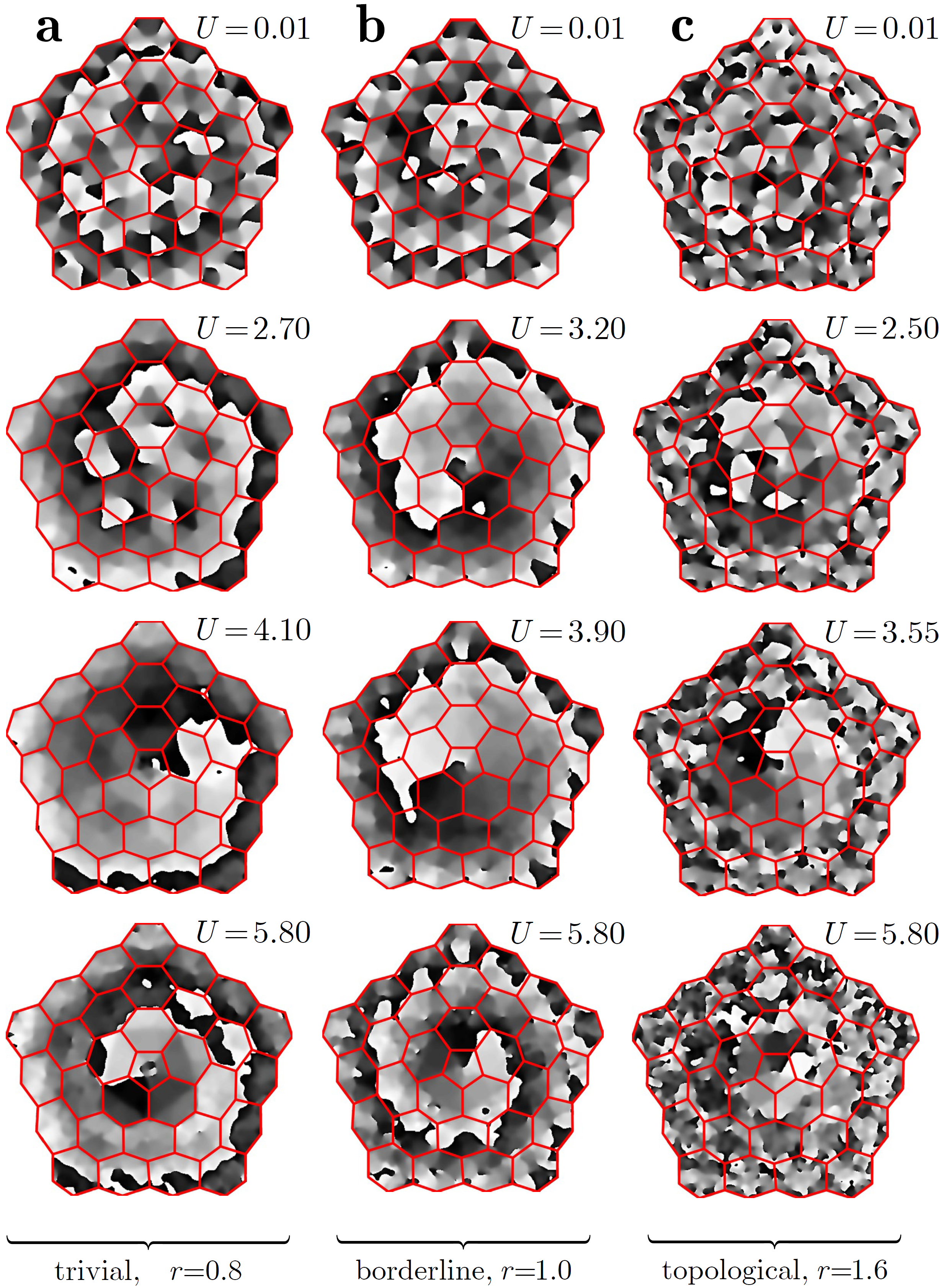}

\justifying
\textbf{Extended Data Fig.~1.} {\bf Excitation of vortex solitons on the disclination core.} Output phase distributions for different input powers corresponding to theoretical output intensity distributions shown in Fig.~3 of the main text. The array is in trivial phase with $r=0.8$ in ({\bf a}), borderline phase with $r=1.0$ in ({\bf b}), and in topological phase with $r=1.6$ in ({\bf c}). Red lines indicate array cells. In color map used here black regions correspond to phase $-\pi$, while white regions correspond to phase $+\pi$.
\label{figED4}
\end{figure*}

\end{document}